 \documentclass[pre,aps,twocolumn,nofootinbib,showpacs,preprintnumbers]{revtex4}
\usepackage{graphicx}% Include figure files
\usepackage{dcolumn}% Align table columns on decimal point
\usepackage{bm}% bold math
\usepackage{amssymb}
\usepackage{pifont}
\usepackage{amsmath}
\usepackage{times}
\usepackage[nooneline]{subfigure}

\begin{document}
\title{Characteristic Lyapunov vectors in chaotic time-delayed systems}

\author{Diego Paz\'o}
\affiliation{Instituto de F\'{\i}sica de Cantabria (IFCA), CSIC--Universidad de
Cantabria, E-39005 Santander, Spain}
\author{Juan M. L{\'o}pez}
\affiliation{Instituto de F\'{\i}sica de Cantabria (IFCA), CSIC--Universidad de
Cantabria, E-39005 Santander, Spain}

\date{\today}

\begin{abstract}
We compute Lyapunov vectors (LVs) corresponding to the largest 
Lyapunov exponents in delay-differential equations with large time delay.
We find that characteristic LVs, and backward (Gram-Schmidt) LVs, 
exhibit long-range correlations, identical to those already observed in dissipative extended systems.
In addition we give numerical and theoretical support to the
hypothesis that the main LV belongs, under a suitable transformation, 
to the universality class of the Kardar-Parisi-Zhang equation. These facts indicate that
in the large delay limit (an important class of)
delayed equations behave exactly as dissipative systems with spatiotemporal chaos.
\end{abstract}

\preprint{submitted to Phys. Rev. E}
\maketitle

\section{Introduction}
\label{sec_intro}
Delayed dynamical systems (DDSs) serve to model diverse phenomena in physics \cite{Erneux} (prominently in optics),
but also in other fields \cite{proc} such as engineering, biology, climatology or ecology.
More than one decade ago \cite{arecchi92,giaco94,giaco96}  it was found that delayed
systems with one constant delay
can be studied like extended systems, and they present not only analogies but equivalent 
phenomena such as pattern instabilities \cite{wolfrum06} and high-dimensional chaos \cite{farmer82}.
Chaos in DDSs is considered as a particular type of spatio-temporal chaos for which the
delay plays the role of the system size.

There are a number of studies concerning the tangent
dynamics of systems with time delay. Previous works
mainly focused on the Lyapunov exponents (LEs) and properties obtained from them
(dimension, entropy, ...)~\cite{farmer82,leberre87,dorizzi87,lepri93,giaco96}. Less is
known about the associated tangent space directions,
in particular, there are no studies analyzing the structure of
Lyapunov vectors (LVs) apart from the main one (pointing along the most unstable direction).
Correlations of the main LV
were only recently addressed~\cite{pik98,sanchez} taking advantage of the
generic mapping of a DDS into an extended system, but with contradicting results.

When considering LVs other than the main one, one must distinguish among different
vector types. The so-called characteristic LVs ~\cite{eckmann} constitute the only intrinsic
set of vectors that is univocally defined and is covariant with the dynamics.
However, their numerical computation is difficult in high-dimensional systems.
We adapt to DDSs one of the methods to compute characteristic LVs \cite{wolfe_tellus07}.

In this paper we report on the generic properties that LVs
exhibit in DDSs in the large delay limit.
Our numerical and theoretical results
indicate that the LVs (the main one and the others) behave in qualitative
and quantitative terms like in one-dimensional dissipative systems with spatiotemporal chaos. 

\section{Time-delayed chaotic systems}
\label{model}
DDSs may exhibit chaos even in the simplest situation in which the main variable is a scalar
and its evolution is determined by a delay-differential equation with one constant delay:
\begin{equation}
\frac{dy}{dt}= {\cal F}(y,y_\tau)
\label{dds}
\end{equation}
where $y_\tau=y(t-\tau)$. In a Lyapunov analysis we are interested in the
evolution of infinitesimal perturbations $\delta y(t)$:
\begin{equation}
\frac{d\delta y}{dt} = u \,  \delta y
+ v \, \delta y_\tau ,
\label{tangent}
\end{equation}
which is also a delayed equation with $\delta y_\tau \equiv \delta y(t-\tau)$, and where
$u$ and $v$ are functions: $u(y,y_\tau)\equiv\partial_y {\cal F}$,
$v(y,y_\tau)\equiv\partial_{y_\tau} {\cal F}$.
If the DDS is chaotic 
an initial random perturbation becomes after some time aligned with the main LV
(reaching a stationary state in a statistical sense),
and the average exponential growth rate is the largest LE of the system $\lambda_1$. Numerically, this is simply achieved
by integrating Eqs.~(\ref{dds}) and (\ref{tangent}) for long enough times.
For non-leading LVs, more involved algorithms are needed (see below).

We focus our study on systems in which delayed and non-delayed terms
are separated: ${\cal F}(y,y_\tau)={\cal Q}(y)+{\cal R}(y_\tau)$.
Our results are expected to be generic for this class of models.
In our study, we have assumed 
the non-delayed part is linear\footnote{We carried out some simulations with ${\cal Q}(y)=-ay^{3/2}$
that did not revealed relevant differences.}, ${\cal Q}(y)=-ay$, and all nonlinearities appear in
the retarded component ${\cal R}$. Many important time-delayed systems,
including the Mackey-Glass (MG)~\cite{mg77} and Ikeda models~\cite{ikeda}, and
optical delayed feedback systems~\cite{goedgebuer98}, can be expressed in
this mathematical form.  In particular, we have studied numerically different nonlinear
functions: the Mackey-Glass (MG), ${\cal R}(\rho)=b \rho /(1+\rho^{10})$,~\cite{mg77};
and the nonlinear function ${\cal R}(\rho)= b \sin^2(\rho-\psi)$ 
that appears in some optical cryptosystems with delayed
feedback~\cite{goedgebuer98,udalstov01}.
Almost identical results in qualitative and quantitative terms are obtained for
these two systems. Therefore, for the sake of brevity, we choose to present
the results only for the MG model, also used in \cite{sanchez}.

In our simulations we have integrated numerically Eq.~(\ref{dds}) using a
third-order Adams-Bashforth-Moulton predictor-corrector
method~\cite{Press}, while
linear equations, {\it e.g.}~(\ref{tangent}),
have been integrated using an Euler method
with the non-delayed part integrated semi-implicitly.
The parameters we used for the MG model are $a=0.1$, $b=0.2$ and the time step
was $dt=0.2$.
In a numerical integration of a time-delayed 
system~\cite{farmer82}, the temporal discretization makes the
system finite-dimensional (with $\tau/dt$ degrees of freedom)
but this has no effect in the results whatsoever since an increase of time
resolution ({\it i.e.}~decreasing $dt$) does not modify the
largest LEs (and their associated vectors).

In several previous works~\cite{arecchi92,giaco94,giaco96,pik98,sanchez} it was found
useful to map the DDS into an equivalent spatially extended
dynamical system of ``size'' $\tau$ that evolves at discrete times $\theta$ as
follows. We express the continuous time as
\begin{equation}
 t=x+\theta \tau
\end{equation}
where $x\in [0,\tau)$ is the ``spatial'' position and $\theta \in \mathbb Z$ is
the discrete time.
This spatial representation can be applied to both, the state of the
system, $y(t) \to y(x,\theta)$, and the perturbations, $\delta y(t) \to
\delta y(x,\theta)$. The benefit of this spatial representation is that one can
analyze the DDS with the tools available for spatially
extended dynamical systems.
We shall use the spatial
representation of delayed systems here to analyze LVs and
analyze the existence of long-range correlations as well as some other dynamical
properties.

\section{Characteristic vs. backward Lyapunov vectors}
\label{sec_lvs}
There is some degree of confusion in the literature regarding the
definition and computation of LVs.
Many authors
think of LVs as the orthonormal frame of vectors that results as a byproduct
of computing the LEs with the algorithm of Benettin et
al.~\cite{benettin80}.
These vectors are usually called Gram-Schmidt or {\em backward}
LVs~\cite{legras96} and they
span the subspaces in tangent space that, at present time $t$, have grown at
exponential rates $\lambda_n$ since the remote past.
Backward LVs $\{b_n(x,\theta)\}$ depend on the particular
scalar product adopted for the Gram-Schmidt orthogonalization in Benettin's
algorithm  (though the subspaces they span are genuine).
This and other undesired properties render
the backward LVs unsuited to analyze problems like
extensivity or hyperbolicity questions.

Ruelle and Eckmann~\cite{ruelle79,eckmann} noticed time ago that one
can define a different set of LVs,
the so-called {\em characteristic} \cite{legras96} LVs $\{g_n(x,\theta)\}$, that are intrinsic
to the dynamics and are univocally defined ({\it i.e.}~independent of how the scalar product is defined).
Each of these vectors is covariant with the dynamics,
and the corresponding $n$-th LE is indeed recovered by either a forward
or backward integration of an infinitesimal perturbation initially
aligned with the $n$-th characteristic LV. Actually, the perturbation will be aligned with the LV 
at all times (hence the covariance).
However for backward LVs (other than the main one, because $b_1=g_1$)
the corresponding LEs are only recovered integrating backward. Until very recently, no efficient
numerical algorithms were available to compute characteristic LVs in large
extended systems.

In this paper we calculate the set of characteristic LVs
by adapting to DDSs the method recently introduced 
by Wolfe and Samelson~\cite{wolfe_tellus07}. In order to
find the $n$th characteristic LV one has to compute the first $n-1$ forward
LVs in addition to the first $n$ backward LVs.
Forward LVs $\{f_n(x,\theta)\}$ are obtained like backward LVs but integrating the perturbations backward in time (from
the remote future to the present time $t$). They have to be calculated 
using the transpose Jacobian matrix so that LEs are obtained with the usual ordering \cite{legras96}: $\lambda_1 \ge \lambda_2
\ge \cdots$. For our time-delayed system this procedure encompasses to
integrate:
\begin{equation}
-\frac{d \delta y(t)}{dt}=u \, \delta y(t) + \tilde v \, \delta y(t+\tau) .
\label{tangentb}
\end{equation}
The diagonal coefficient
$u = \left.\partial_{w} {\cal F}(w,z)\right|_{w=y(t), z=y(t-\tau)}$
is identical to that in Eq.~(\ref{tangent}).
The off-diagonal term has a curious structure stemming from
transposing a Jacobian matrix with nonzero elements $\tau$ temporal units
below the main diagonal.
Thus, in Eq.~(\ref{tangentb}), we have  $\tilde v =\left.\partial_{z} {\cal F}(w,z)\right|_{w=y(t+\tau), z=y(t)}$.
Notice that,
as the integration runs backward in time ($t\to-\infty$), the minus sign in the
left hand side is canceled.
As occurs with backward LVs, periodic Gram-Schmidt orthonormalizations are needed
to avoid the collapse of all perturbations along the most unstable direction.
Note also that, due to the delay, Eq.~(\ref{dds}) cannot be integrated backward,
and hence it is necessary to store the trajectory $y(t)$ at every time step in
the computer to be used in the computation of~(\ref{tangentb}).

Once both backward LVs and forward LVs have been computed the characteristic LVs
are easily calculated following the prescriptions in Ref.~\cite{wolfe_tellus07}. One delicate point
in delayed systems is that one must make sure that both forward and backward LV sets 
correspond exactly to the same time interval $[t,t+\tau)$.

\section{Long-range correlations of Lyapunov vectors}
\label{sec_struct}
 In this section we study the form of both backward
(Gram-Schmidt) and characteristic LVs corresponding to the largest LEs,
their localization properties, and the existence of long-range correlations.

LVs are strong localized objects, therefore, 
like in Refs.~\cite{pik98,sanchez}, it is very convenient
to work with the associated fields obtained after a logarithmic
transformation:
\begin{equation}
h_n(x,\theta)= \ln |g_n(x,\theta)|
\end{equation}
(and likewise for backward LVs).
We will refer to $h_n$ as the $n$th {\em surface} due to the similarities of its
dynamics with that of the kinetic roughening of fractal surfaces in growth models
(see details below). Also recall that the norm is irrelevant
because a LV only indicates a direction in tangent space, and this implies
that the mean height of $h_n$ is irrelevant (only its profile fluctuations matter);
this should be beard in mind for the theoretical analysis below.

%%%%%%%%%%%%%%%%%%%%%%%%%%%%%%%%%%%%%%%%%%%%%%%%%%%%%%%%%%%%%%%%%%%%%%
\begin{figure}
\centerline{\includegraphics *[width=75mm]{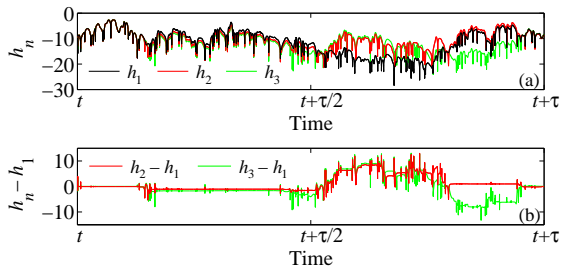}}
\caption{(Color online) (a) Snapshot of the surfaces corresponding to
characteristic LVs $n=1,2,3$ ($\tau=3277$ t.u.).
(b) Difference of 2nd and 3rd LV-surfaces with respect to the 1st LV-surface.
The existence of plateaus evidences the piecewise copy
structure of LV surfaces with respect to the 1st one.}
\label{vectors}
\end{figure}
%%%%%%%%%%%%%%%%%%%%%%%%%%%%%%%%%%%%%%%%%%%%%%%%%%%%%%%%%%%%%%%%%%%%%%

Figure~\ref{vectors}(a) shows a snapshot of the characteristic LV surfaces
$h_1$, $h_2$, and $h_3$
corresponding to
the three largest LEs\footnote{For $\tau=3277$ t.u.~there are 168 positive LEs.
The $n_0$-th LE vanishes, with $n_0\simeq0.0514 \, \tau+0.79$.}. We find that non-leading LV surfaces are approximately
piecewise copies of the leading LV surface $h_1$, see Fig.~\ref{vectors}(b). This
`replication property' 
is a highly nontrivial phenomenon that was originally discovered to occur in
chaotic extended dissipative systems~\cite{szendro07,pazo08}.
We also emphasize that characteristic LVs exhibit
a tendency to clusterize, contrary to backward LVs whose localization sites are
scattered due to the imposed orthogonality. Indeed, as exemplified by the
snapshot shown in Fig.~\ref{vectors}(a),
the first three characteristic LVs localize (reach their largest magnitude)
at the same point
(completely overlapping on the leftmost region in this particular snapshot).
This does not occur all the time but in an intermittent manner.

Interestingly, these structural features-- namely, replication and
clustering-- have recently been reported to
occur
generically for LVs in chaotic spatially extended 
systems~\cite{szendro07,pazo08,mauricio}. This
deepens in the analogy between DDSs and systems with extensive
chaos in one dimension, which happens to hold even at the level of non-leading LVs.

Regarding the existence of long-range correlations,
we find that LV surfaces have a self-affine spatial structure at long scales, which
translates into power-law correlations.
A detailed analysis of the spatial structure can be best achieved by
computing the Fourier transform of the surfaces at discrete times $\theta$ given
by
$\hat h_n(k,\theta)= \frac{1}{\sqrt{\tau}} \int_0^{\tau} \exp(ikx) \,
h_n(x,\theta) \, dx$,
with wavenumbers $k\in[\tfrac{2\pi}{\tau},\tfrac{\pi}{dt}]$, where $\tau$ is the
system size in this representation. For a given LV surface $h_n$, the Fourier
transform of the two-point correlator $\langle h_n(x_0+x,\theta)
h_n(x_0,\theta)\rangle - \langle h_n(x_0,\theta)\rangle^2$, is the so-called structure
factor:
\begin{equation}
 S_n(k)=\langle \hat h_n(k,\theta) \hat h_n(-k,\theta) \rangle
\end{equation}
where the brackets denote average over time $\theta$ and realizations. Therefore, the
structure
factor $S_n(k)$ directly informs about the $n$th LV surface correlations
at scale $1/k$.

%%%%%%%%%%%%%%%%%%%%%%%%%%%%%%%%%%%%%%%%%%%%%%%%%%%%%%%%%%%%%%%%%%%%%%
\begin{figure}
\centerline{\includegraphics *[width=75mm]{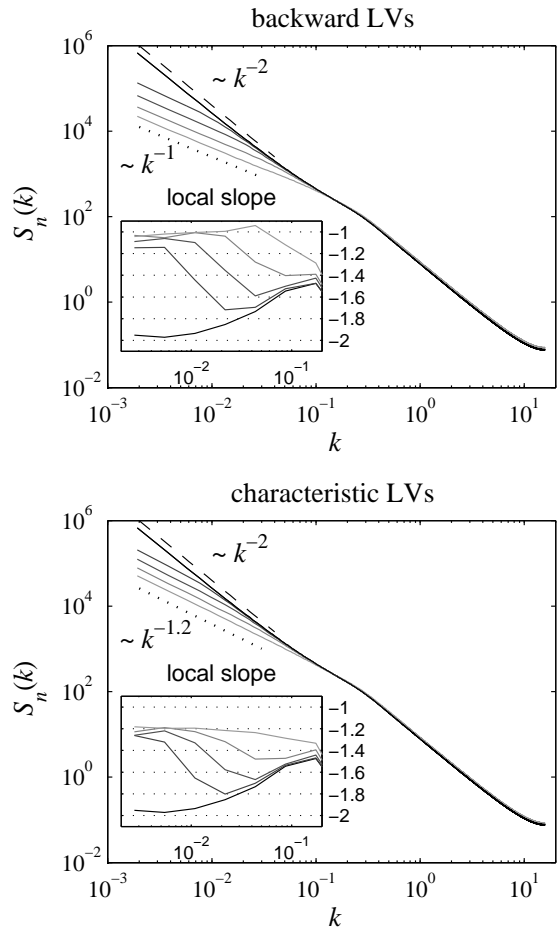}}
\caption{Structure factors of LV-surfaces for the MG model with $\tau=3277$ t.u.
The curves from top to bottom correspond to $n=1,4,8,16,32$. The insets show the
dependence of the local
slopes of different curves. For $n=1$ the asymptotic slope is $-2$ whereas for
$n>1$ the asymptotic slopes
are $\approx -1$ and $\approx -1.2$ for backward and characteristic LVs,
respectively.}
\label{sf}
\end{figure}
%%%%%%%%%%%%%%%%%%%%%%%%%%%%%%%%%%%%%%%%%%%%%%%%%%%%%%%%%%%%%%%%%%%%%%

In Fig.~\ref{sf} it can be seen that the structure factor for the first
LV-surface decays asymptotically as $k^{-(2\alpha+1)}$, with the so-called roughness exponent
$\alpha=1/2$ (in agreement with the result in \cite{pik98} for the Ikeda model).
Figure \ref{sf} also reveals the existence of 
a particular crossover wavenumber $k_n^\times\approx
\tfrac{2\pi n}{\tau}$ for the $n$th LV.
The exponents observed for $k<k_n^\times$ are $-1.2$ and $-1$ for characteristic and backward LVs,
respectively. These numbers coincide with those previously reported in dissipative
systems in one dimension~\cite{szendro07,pazo08}.
Our interpretation of these results is that for backward LVs the $k^{-1}$
dependence seems to be the consequence of residual correlations with a geometric origin
in the orthogonality of the basis.
The $-1.2$ exponent of characteristic LVs indicates they convey more information among
distant parts of the system than backward LVs; which can be understood as a
consequence of the covariance of characteristic LVs as they are consistent with both past and future
evolutions [characteristic LVs with $n>1$ are ``saddle solutions'' of
(\ref{tangent}) forced by (\ref{dds}), see \cite{pazo08}].

\section{Main {L}yapunov vector and the universality class question}
As we have seen in the previous section, the surface $h_1$ associated with the main
LV exhibits scale-invariant correlations in the $k\to 0$ limit. This strongly
suggests that this LV surface should belong to one of the
universality classes of surface growth. The fact that the LV surfaces
obey scaling laws and that systems with spatiotemporal chaos could be divided
into a few universality classes, according to the scaling of the associated
surfaces, has implications in our understanding of chaos in extended
systems from a statistical physics point of view.
Moreover, the generic replication and clustering of characteristic LVs along the
main vector direction in extended
systems~\cite{szendro07,pazo08,mauricio} (also
observed here for time-delay systems, see previous section) makes it even more
interesting to
determine the universality class of the main LV, since this is expected to
provide a great deal of information about the structure of space and time
correlations of the LV corresponding to leading as well as sub-leading unstable
directions.

For a wide class of one-dimensional spatio-temporal chaotic systems (see~\cite{pik98}) the
statistical features of the main LV surface are well captured by the
1$d$-Kardar-Parisi-Zhang (KPZ)~\cite{kpz}
stochastic surface growth equation:
\begin{equation}
\partial_t h(x,t) = 
\chi(x,t) + \kappa (\partial_x h)^2 + \nu \partial_{xx} h ,
\label{kpz_eq}
\end{equation}
where $\chi$ is a white noise.
The KPZ equation defines itself an important universality class of
surface growth.
Regarding chaotic systems, in many of them the main LV surface belongs to the KPZ class.
This includes coupled-map lattices, coupled symplectic maps, the complex Ginzburg-Landau
model, Lorenz 96 model, among others~\cite{pik94,pik98,szendro07,pazo08}. Only LVs of anharmonic Hamiltonian lattices
are known~\cite{pik01} to exhibit correlations that are clearly inconsistent
with KPZ exponents ($\alpha>\alpha_{KPZ}=1/2$).

Statistical mechanics teaches us that, given the fact that KPZ represents a
dynamical universality class, one would expect that a large collection of
different systems could belong to the KPZ universality despite their apparent
differences. Only symmetries and conservation laws would determine the
universality class. That may explain why Hamiltonian (energy conserving) systems,
in contrast to dissipative systems, do not generally belong to the KPZ universality.
On this basis, DDSs were also proposed~\cite{sanchez} to belong to a different
universality class, namely the Zhang model, arguing that these systems break the $x \to -x$ symmetry
in the spatial representation. However, this conclusion was in contradiction
with an earlier work~\cite{pik98} where KPZ was postulated.
Our aim in this section is to clarify this question by means
of a theoretical analysis and extensive numerical simulations.

\subsection{Theoretical analysis: KPZ vs.~Zhang equation}
Our starting point is the evolution equation in
tangent space (\ref{tangent}), of which the main LV is the asymptotic solution.
In the spatial representation (see Sec.~\ref{model}) we
perform the change of variables $\theta = (t-x)/\tau$, so that (\ref{tangent})
is now written as:
\begin{equation}
\partial_x \delta y(x,\theta) = u \, \delta y(x,\theta) + v \, \delta y(x,\theta-1),
\label{2}
\end{equation}
where $x\in [0,\tau)$ is the `spatial' position and $\theta \in \mathbb Z$ is
the discrete time.
We have seen in the previous section that a description in terms of surfaces instead
of vectors themselves is more appropriate due to the strong localization
of the latter. Moreover, to relate Eq.~(\ref{2}) with one of the stochastic
partial differential equations modeling surface growth,
we wish to approximate the discrete differences in $\theta$
by a partial derivative. In sum, we have to take two steps:

\begin{enumerate}
\item[(i)] Transform to a surface: $h(x,\theta)=\ln |\delta y(x,\theta)|$.
\item[(ii)] Approximate $\theta$ by a continuous variable:  $s(\theta)-s(\theta-1)\to \partial_\theta s$.
\end{enumerate}
Note that these two steps do not commute.

\subsubsection{Linear Zhang model}
\label{zhang_sec}
The treatment of Eq.~(\ref{2}) by S\'anchez {\it et. al.}~\cite{sanchez}
proceeded with step (ii) before step (i).
In more detail, after step (ii), Eq.~(\ref{2}) becomes:
\begin{equation}
\partial_\theta \delta y= -(1/v) \, \partial_x \delta y + (u/v + 1) \delta y ,
\end{equation}
and now transforming to the surface picture [step (i)] one has:
\begin{equation}
 \partial_\theta h(x,\theta) = - (1/v) \partial_x h + (u/v) +1,
\label{h_sanchez}
\end{equation}
so that we get an equation for the field $h(x,\theta)$. This equation was already analyzed
in Ref.~\cite{sanchez} and we summarize some of its properties in the following.
On the one hand, the random drift term $(1/v) \partial_x h$ gives rise to
an effective diffusion term $D \partial_{xx} h$. On the other hand,
the presence of $v$ in the denominator 
induces large fluctuations of those terms
proportional to $\zeta=1/v$ every time that $v$ takes values close to zero.
Usually, the probability density 
function is algebraic at the lowest order:
\begin{equation}
 P(v) \sim |v|^\sigma \qquad (|v|\ll 1),
\label{pv}
\end{equation}
with $\sigma>-1$ to be normalizable (and $\sigma=0$ in general).
In turn, $P(\zeta)$ is heavy tailed with large fluctuations $P(\zeta)\sim
|\zeta|^{-(2+\sigma)}$
(this is indeed in agreement with data we have collected in our simulations of
the MG model: $\sigma\approx0$).
These reasoning led the authors of Ref.~\cite{sanchez} to conclude that
the main LV surface in DDSs behaves following (\ref{h_sanchez}) and
generically falls into the universality class of the linear Zhang model~\cite{zhang90}:
\begin{equation}
\partial_\theta h(x,\theta) =  D \partial_{xx} h + c + \xi(x,\theta),
\label{zhang_eq}
\end{equation}
that describes surface growth driven by a heavy-tailed noise $\xi(x,\theta)$
with a probability distribution $P(\xi) \sim |\xi|^{-(1+\mu)}$ for $|\xi| \gg 1$ and
the index $\mu=2(1+\sigma)$ ($\ge 2$ if $\sigma\ge 0$).

\subsubsection{KPZ universality}
In this work we propose an alternative analysis of Eq.~(\ref{2}),
where step (i) is carried out before step (ii). This would
be more suited because one applies the nonlinear transformation (i)
prior to the approximation (ii).
Thus after defining the corresponding LV surface, $h(x,\theta)=\ln |\delta y (x,\theta)|$, 
Eq.~(\ref{2}) becomes
\begin{equation}
 \partial_x h(x,\theta) = 
u \pm v \exp{[h(x,\theta-1)-h(x,\theta)]},
\label{interm}
\end{equation}
where the choice of sign $\pm$ comes from the absolute value and is irrelevant
for the arguments that follow.
We approximate the difference $h(x,\theta-1)-h(x,\theta)$
by the partial derivative $-\partial_\theta h$.
With no further approximations\footnote{Considering the time derivative $\partial_\theta h$
small enough so that the exponential in Eq.~(\ref{interm}) can be expanded and
only the lowest order may be retained yields Eq.~(\ref{h_sanchez}).
We propose here a different derivation that avoids this
assumption and, remarkably, leads to a different result.}
and taking logarithms in Eq.~(\ref{interm}) we get:
\begin{equation}
\partial_\theta h(x,\theta) = \ln|v| - \ln|\partial_x h - u| 
\label{logeq}
\end{equation}
where the term $\eta=\ln|v|$ is again a fluctuating noise-like source due to
the chaotic character of the trajectories.
In the neighborhood of $v=0$ the probability distribution of $v$ is (\ref{pv}), and thus
large values of $|\eta|$ are exponentially rare:
$P(\eta\to -\infty)\sim e^{-(\sigma+1)|\eta|}$, {\it i.e.}~the noise is not heavy tailed.

We can obtain a more intuitive equation by making use of
the small gradient approximation, $|\partial_x h| \ll 1$.
Expanding~(\ref{logeq}) in the form:
\begin{equation}
\partial_\theta h(x,\theta) = \eta - \ln|u|+\frac{\partial_x h}{u} + 
\frac{(\partial_x h)^2}{2u^2} + O[(\partial_x h)^3] , 
\label{logeq2}
\end{equation}
where the prototypical quadratic term $(\partial_x h)^2$ of KPZ appears.
If $u$ is fluctuating, the drift term
$(\partial_x h)/u$ again yields (at large scales, after a spatial averaging)
an effective diffusion term. In this case, Eq.~(\ref{logeq2})
would lead to the KPZ Eq.~(\ref{kpz_eq}).
In contrast, if $u$ is a constant, like for instance in the case of the MG model,
a diffusion term would also eventually appear
far from the small gradient limit as a result of the absolute value 
in $\ln|\partial_x h - u|$ in~(\ref{logeq}).
Finally, notice that the noise-like term $\ln(|v|/|u|)$, which is different from that
in~(\ref{zhang_eq}), will not lead to rare events and
its distribution will be exponentially decaying in general.

At variance with S\'anchez {\it et al.}~derivation~\cite{sanchez}
({\it cf.}~Sec.~\ref{zhang_sec}) we have arrived here at the LV surface
equation~(\ref{logeq2})
by expanding on the local slope $\partial_x h$, which is generally expected to
be a reasonable approach when the roughness exponent $\alpha < 1$ since the
spatial average over a window of extent $\ell$ should scale as $\langle
|\partial_x h| \rangle_\ell \sim \ell^{\alpha - 1}$ and goes to zero as we
coarse-grain $\ell \gg 1$.

%%%%%%%%%%%%%%%%%%%%%%%%%%%%%%%%%%%%%%%%%%%%%%%%%%%%%%%%%%%%%%%%%%%%%%
\begin{figure}
\centerline{\includegraphics *[width=75mm]{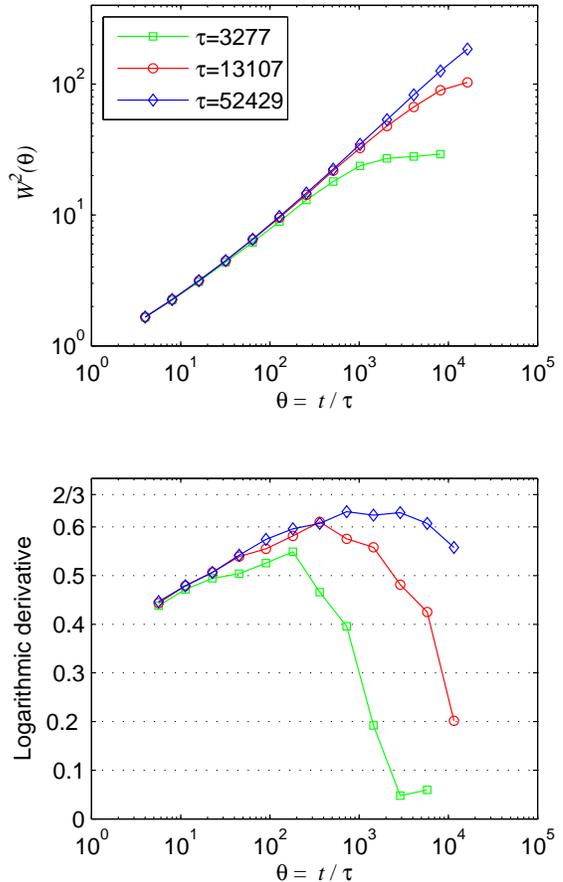}}
\caption{(Color online) Evolution of the width from a random perturbation for
three different
values of $\tau$.
The lower panel shows the local derivative.
For the largest value of the delay, $\tau=52429$ t.u., the local derivative
exhibits
a plateau about $0.63$, close to the theoretical value for KPZ ($2\beta=2/3$). 
The curves are averages over at least 1000 realizations.}
\label{beta}
\end{figure}
%%%%%%%%%%%%%%%%%%%%%%%%%%%%%%%%%%%%%%%%%%%%%%%%%%%%%%%%%%%%%%%%%%%%%%

\subsection{Growth exponent}
In the field of growing interfaces it is customary to quantify the temporal
features by looking at the growth of the surface width at a given time when started
from a flat initial perturbation [$h(x,0)\approx \mathrm{const.}$]. To do so we let
random initial perturbations $\delta y(x,0)$ to evolve and measure the
average surface width at time $\theta$ as
\begin{equation}
W^2(\theta) = \left\langle \overline{\left[h(x,\theta)-
\overline{h}(\theta) \right]^2} \right\rangle,
\end{equation}
where $\overline{h}(\theta)= \tau^{-1}\int_{0}^\tau h(x,\theta) dx$ is the mean
surface position and brackets denote averaging over
different initial realizations.
The growth exponent $\beta$ is defined as the exponent of the transient
power-law growth: $W^2 \sim \theta^{2\beta}$,
before the surface fluctuations saturate (for $\theta\ll\theta_\times$); while
$W^2 \sim \tau^{2\alpha}$, when saturation is reached (for $\theta\gg\theta_\times$) 
and the perturbation is virtually aligned with the LV: $\delta y \propto g_1 \Rightarrow h=h_1+{\rm const}$. The
key point now is that $\beta$
equals $1/3$ for KPZ, while it is $1/4$ for the Zhang model with $\mu \ge
2$~\cite{lam93}. This makes the time exponent $\beta$ an excellent index
to distinguish between KPZ and Zhang behavior
(note that the spatial exponent $\alpha$ is the same in both models: $\alpha=1/2$).

Figure~\ref{beta} shows that $W^2$ grows with an exponent that, for
large enough delays, progressively approaches the KPZ growth exponent
$\beta_{KPZ}=1/3$, which strongly supports the KPZ asymptotic ($\tau \to
\infty$) scaling for chaotic DDSs.
In the paper by S\'anchez et al.~\cite{sanchez} the largest time delay used 
$\tau=4000$ t.u.~resulted in an estimation $\beta=1/4$. In the light of our simulations
with much larger delays we conclude that the
value of $\tau$ used in Ref.~\cite{sanchez} was insufficient to detect
the true asymptotic scaling exponent.

\subsection{Multiscaling}

In addition to the growth exponent, Ref.~\cite{sanchez} also invoked the existence
of multiscaling of the main LV surface as an argument supporting its identification with the
universality class of the linear Zhang model (\ref{zhang_eq}).
This model, contrary to the KPZ equation, leads to surfaces that exhibit
multiscaling, {\it i.e.}~strongly non-Gaussian tails,
induced by extreme events.
If we compute the
$q$th-height-height correlation function for points separated a distance $l$,
\begin{equation}
\label{G_q}
G_q(l)=\left\langle \overline{|h_1(x+l,\theta)-h_1(x,\theta)|^q}
\right\rangle^{1/q},
\end{equation}
(where the overline denotes the average over $x$ and the
brackets denote a realizations average)
one finds that $G_q(l) \sim l^{\alpha_q}$, and multiscaling exists
if the roughness exponents $\alpha_q$ depend on the index $q$. 

%%%%%%%%%%%%%%%%%%%%%%%%%%%%%%%%%%%%%%%%%%%%%%%%%%%%%%%%%%%%%%%%%%%%%
\begin{figure}
\centerline{\includegraphics*[width=75mm]{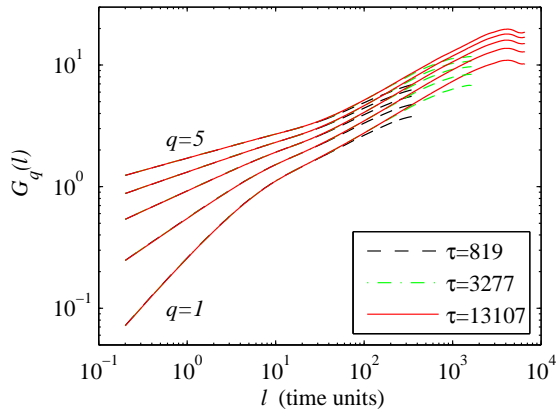}}
\caption{(Color online) Height-height correlation functions $G_q(l)$,
$q=1,\ldots,5$.
Multiscaling is observed at small $l$. Above a certain characteristic size
$l_c\sim 10$
no multiscaling is observed. Note that $l_c$ is insensitive to $\tau$, and the
curves overlap
almost perfectly (except at large $l$ due to finite-size effects).}
\label{multiscaling}
\end{figure}
%%%%%%%%%%%%%%%%%%%%%%%%%%%%%%%%%%%%%%%%%%%%%%%%%%%%%%%%%%%%%%%%%%%%%%

We have computed the $q$th order height-height correlation function $G_q(l)$,
given by Eq.~(\ref{G_q}), for several values of $\tau$ ({\it i.e.}~system sizes);
see Fig.~\ref{multiscaling}. The distribution of the
surface fluctuations shows clear signs of multiscaling with local roughness
exponents $\alpha_q$ that depend on $q$ for length scales below a typical
scale of about $l_c \sim 10$. These non-Gaussian features disappear at larger
scales ($l \gg l_c$), where the $\alpha_q \approx 1/2$ for all $q$.
For truly extreme
event dominated fluctuations, this length scale is expected to diverge (however
slowly) with the system size, so that the fluctuation distribution is truly
non-Gaussian in the thermodynamic limit. This slow divergence was indeed
measured for models in the Zhang universality class~\cite{barabasi92}. In the
case of time-delay systems, this would correspond to having $l_c(\tau)$
increasing with $\tau$. In contrast, we find that the characteristic length
$l_c\sim 10$ remains constant even after an increase of the delay of 16 times.
Again, the simulations by S\'anchez {\it et. al.}~\cite{sanchez} were carried out in
systems with delays that were too short to obtain conclusive evidence on the dependence of
$l_c$ with $\tau$. Certainly, we cannot rule out for sure an extremely weak
dependence of $l_c$ on $\tau$, but if it exists it must be sub-logarithmic and
well beyond the precision that we can reach in our simulations. We conclude
that multiscaling of the LV surface fluctuations in DDSs seems to
be a short scale phenomenon that has no effect in the thermodynamic limit where
the universality class is defined; in this case described by the KPZ equation.

\section{Conclusions}

In this work we have implemented (to our knowledge for the first time) characteristic
LVs in DDSs. Adaptation of the method proposed in \cite{wolfe_tellus07}
to this kind of systems ---together with the computer capabilities nowadays available---
allowed us to reach systems with fairly large delays, which serves to investigate the
``thermodinamic limit'' of these systems. Our results for the LVs coincide
quantitatively with those obtained in extended dissipative systems~\cite{szendro07,pazo08}.

In addition we have revisited the question of which universality class the main LV
belongs to. After simulations with very large delays we may conclude that the main LV surface
falls into the universality class of the Kardar-Parisi-Zhang equation.
Our theoretical arguments support this conclusion as well.

In sum, DDS are equivalent to extended dynamical systems in the sense that
infinitesimal perturbations exhibit the same exponents characterizing
spatiotemporal correlations.
DDS have been traditionally considered to be
different because of the lack of extensivity of the Lyapunov spectrum:
the positive exponents approach zero as $\sim 1/\tau$ \cite{farmer82}
and the (Kolmogorov-Sinai) entropy saturates with $\tau$. However the identification
of $\tau$ with a size implies that comparisons
should be done in temporal units of $\theta=t/\tau$, and extensivity is then recovered.
As $\lambda_n t= \lambda_n \tau \theta= \Lambda_n \theta$,
the redefined LEs $\Lambda_n=\tau\lambda_n$  do not decay to zero with $\tau$,
and the Lyapunov spectrum converges to a density in the thermodynamic limit.

\acknowledgments
D.P.~acknowledges support by CSIC under the Junta de Ampliaci\'on de
Estudios Programme (JAE-Doc). Financial support from
the Ministerio de Ciencia e Innovaci\'on (Spain) under project
No.~FIS2009-12964-C05-05 is acknowledged.


\begin{thebibliography}{10}

\bibitem{Erneux}
T. Erneux, {\em Applied Delay Differential Equations} (Springer, New York,
  2009).

\bibitem{proc} {\em Delayed complex systems} in {\em Phil. Trans. R. Soc. A} {\bf 368}, No.~1911,
edited by W. Just, A.~Pelster, M. Schanz, and E. Sch\"oll (2010).

\bibitem{arecchi92}
F.~T. Arecchi, G. Giacomelli, A. Lapucci, and R. Meucci, Phys. Rev. A {\bf 45},
   R4225  (1992).

\bibitem{giaco94}
G. Giacomelli, R. Meucci, A. Politi, and F.~T. Arecchi, Phys. Rev. Lett. {\bf
  73},  1099  (1994).

\bibitem{giaco96}
G. Giacomelli and A. Politi, Phys. Rev. Lett. {\bf 76},  2686  (1996).

\bibitem{wolfrum06}
M. Wolfrum and S. Yanchuk, Phys. Rev. Lett. {\bf 96},  220201  (2006).

\bibitem{farmer82}
J.~D. Farmer, Physica D {\bf 4},  366  (1982).

\bibitem{leberre87}
M. Le~Berre {\it et~al.}, Phys. Rev. A {\bf 35},  4020  (1987).

\bibitem{dorizzi87}
B. Dorizzi {\it et~al.}, Phys. Rev. A {\bf 35},  328  (1987).

\bibitem{lepri93}
S. Lepri, G. Giacomelli, A. Politi, and F.~T. Arecchi, Physica D {\bf 70},  235
   (1993).

\bibitem{sanchez}
A.~D. S\'anchez, J.~M. L{\'o}pez, M.~A. Rodr{\'\i}guez, and M.~A. Mat\'{\i}as,
  Phys. Rev. Lett. {\bf 92},  204101  (2004).

\bibitem{pik98}
A. Pikovsky and A. Politi, Nonlinearity {\bf 11},  1049  (1998).

\bibitem{eckmann}
J.-P. Eckmann and D. Ruelle, Rev. Mod. Phys. {\bf 57},  617  (1985).

\bibitem{wolfe_tellus07}
C.~L. Wolfe and R.~M. Samelson, Tellus {\bf 59A},  355  (2007).

\bibitem{mg77}
M.~C. Mackey and L. Glass, Science {\bf 197},  287  (1977).

\bibitem{ikeda}
K. Ikeda, Opt. Commun. {\bf 30},  257  (1979).

\bibitem{goedgebuer98}
J.-P. Goedgebuer, L. Larger, and H. Porte, Phys. Rev. Lett. {\bf 80},  2249
  (1998).

\bibitem{udalstov01}
V.~S. Udaltsov, J.-P. Goedgebuer, L. Larger, and W.~T. Rhodes, Phys. Rev. Lett.
  {\bf 86},  1892  (2001).

\bibitem{Press}
W.~H. Press, S.~A. Teukolsky, W.~T. Vetterling, and B.~P. Flannery, {\em
  {N}umerical {R}ecipes in {F}ortran 77: {T}he {A}rt of {S}cientific
  {C}omputing}, 2nd ed. ({C}ambridge {U}niversity {P}ress, Cambridge, 1992).

\bibitem{benettin80}
G. Benettin, L. Galgani, A. Giorgilli, and J.-M. Strelcyn, Meccanica {\bf 15},
  9  (1980).

\bibitem{legras96}
B. Legras and R. Vautard,  in {\em Proc. Seminar on Predictability Vol. I},
  ECWF Seminar, edited by T. Palmer (ECMWF, Reading, UK, 1996), pp.\ 135--146.

\bibitem{ruelle79}
D. Ruelle, Publ. Math. IHES {\bf 50},  27  (1979).

\bibitem{szendro07}
I.~G. Szendro, D. Paz{\'o}, M.~A. Rodr{\'\i}guez, and J.~M. L{\'o}pez, Phys.
  Rev. E {\bf 76},  025202(R)  (2007).

\bibitem{pazo08}
D. Paz\'{o}, I.~G. Szendro, J.~M. L\'{o}pez, and M.~A. Rodr\'{\i}guez, Phys.
  Rev. E {\bf 78},  016209  (2008).

\bibitem{mauricio}
M. Romero-Bastida, D. Paz\'{o}, J.~M. L\'{o}pez, and M.~A. Rodr\'{\i}guez, Phys.
  Rev. E {\bf 82},  036205  (2010).

\bibitem{barabasi92}
A.-L. Barab\'asi {\it et~al.}, Phys. Rev. A {\bf 45},  R6951  (1992).

\bibitem{kpz}
M. Kardar, G. Parisi, and Y.-C. Zhang, Phys. Rev. Lett. {\bf 56},  889  (1986).

\bibitem{pik94}
A.~S. Pikovsky and J. Kurths, Phys. Rev. E {\bf 49},  898  (1994).

\bibitem{pik01}
A. Pikovsky and A. Politi, Phys. Rev. E {\bf 63},  036207  (2001).

\bibitem{zhang90}
Y.-C. Zhang, J. Phys. (France) {\bf 51},  2129  (1990).

\bibitem{lam93}
C.-H. Lam and L.~M. Sander, Phys. Rev. E {\bf 48},  979  (1993).

\end{thebibliography}
\end{document}